\pdfoutput=1
\documentclass[a4paper,aip]{revtex4-1}
\usepackage[utf8]{inputenc}
\usepackage{amsmath}
\usepackage{amsfonts}
\usepackage{amssymb}
\usepackage{bm}
\usepackage{graphicx}
\usepackage{dcolumn}
\usepackage{natbib}
\usepackage{units}
\usepackage[T1]{fontenc}
\usepackage{mathptmx}

\begin{document}

\author{C. Arran}
\thanks{christopher.arran@york.ac.uk}
\affiliation{York Plasma Institute, Department of Physics, University of York, YO10 5DD}
\author{C. P. Ridgers}
\affiliation{York Plasma Institute, Department of Physics, University of York, YO10 5DD}
\author{N. Woolsey}
\affiliation{York Plasma Institute, Department of Physics, University of York, YO10 5DD}

\title{Proton Radiography in Background Magnetic Fields}

\begin{abstract}
Proton radiography has proved increasingly successful as a diagnostic for electric and magnetic fields in high energy density physics experiments. Most experiments use target-normal-sheath-acceleration sources with a wide energy range in the proton beam, as the velocity spread can help differentiate between electric and magnetic fields and provide time histories in a single shot. However, in magnetised plasma experiments with strong background fields, the broadband proton spectrum leads to velocity-spread-dependent displacement of the beam and significant blurring of the radiograph. We describe the origins of this blurring and show how it can be removed from the experimental measurement, and we outline the conditions under which such deconvolutions are successful. As an example, we apply this method to a magnetised plasma experiment that used a background magnetic field of $\unit[3]{T}$. The strong displacement and energy spread of the proton beam reduced the spatial resolution from tens of microns to a few millimetres. The deconvolution procedure is applied showing the accurate recovery of radiographs with resolutions better than $\unit[100]{\mu m}$, enabling the recovery of more accurate estimates of the path integrated magnetic field. This work extends accurate proton radiography to a class of experiments with significant background magnetic fields, particularly those experiments with an applied external magnetic field.
\end{abstract}

\maketitle

\section{Introduction}


Laser-driven proton radiography has proven an essential diagnostic for measuring magnetic field structures in plasmas. Typically, protons are produced using target-normal sheath acceleration (TNSA) \cite{borghesi2006}, pass through a magnetic field region of interest, and are measured by the proton dose absorbed by a stack of radiochromic film (RCF). By measuring the intensity pattern of the radiographs from different proton energies on different layers of film, the structure of the path-integrated magnetic field in both space and time can be accurately recovered \cite{kugland2012, kasim2019}, with high spatial resolution and laser synchronisation providing excellent time resolution. This technique has been used to study a host of effects from Nernst advection \cite{willingale2010, li2013} to magnetic reconnection \cite{palmer2019, tubman2021} and is of vital importance for studies of magnetised high energy density physics. As magnetic fields in plasmas are of great interest both in laboratory astrophysics experiments \cite{gregori2012, meinecke2014}, and for suppressing heat flow and instability growth and enhancing yield in inertial confinement fusion experiments \cite{perkins2017, walsh2019} and hybrid magnetised fusion schemes \cite{slutz2012, sefkow2014}, it seems likely that proton radiography is only going to become more useful with time.

Several studies have sought to use an applied background magnetic field to explore conditions in a magnetised plasma \cite{chang2011,hohenberger2012}, and platforms to apply and measure strong pulsed-power magnetic fields to a plasma target are being developed at a number of laser facilities \cite{manuel2015,albertazzi2018,bradford2020}. Under these conditions, however, the performance of proton radiography can be severely affected due to the deflection of the proton beam in the background fields, which can overwhelm the signal from the magnetic field inside the plasma. Furthermore, the deflection of protons in the background field is energy dependent, and this introduces problems for proton radiography with broadband sources.

In this paper, we seek to understand and remedy the effects of proton deflection in a strong background field. A recent experiment used laser-driven proton radiography to measure changes to an applied magnetic field. In the process we observed both substantial deflection of the proton beam and also a blurring effect, where the spatial resolution of the radiographs was dramatically reduced in the direction of the deflection. Firstly, we show that this issue originates from using a broadband energy spectrum TNSA proton source in combination with a stack of radiochromic film which absorbs protons over a finite range of energies. Secondly, by understanding the source of the blurring, we show how this effect can be modelled by a linear convolution. Finally, we explore how a deconvolution process can recover a more accurate estimate for the magnetic field profile, and we demonstrate the conditions under which this deconvolution is successful and its limitations.

\section{Blurring in a background field}


\begin{figure*}
\begin{center}
\includegraphics[scale=0.9]{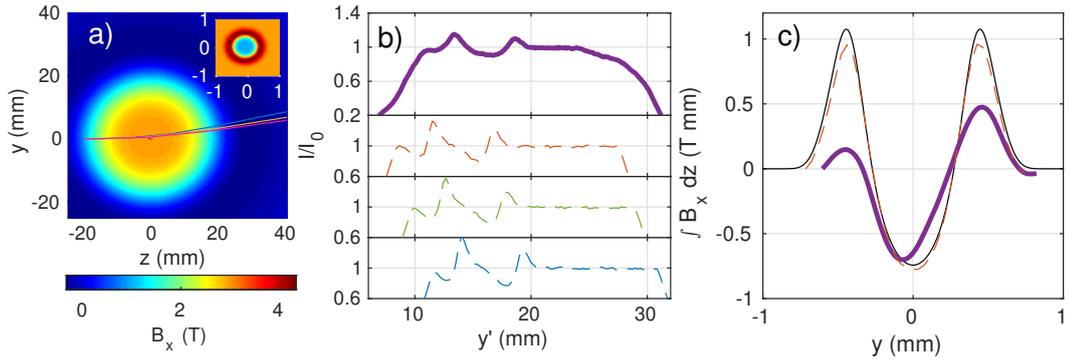}
\caption{a) An example of the magnetic field structure, with a large background field surrounding a mm-scale signal region near the centre (close-up shown in the inset on the top right). Overlaid are the paths of mono-energetic proton beams of energies $\unit[10]{MeV}$ (blue), $\unit[16]{MeV}$ (yellow), and $\unit[21]{MeV}$ (pink). b) An example of three synthetic mono-energetic radiographs (dashed lines) at $\unit[10]{MeV}$ (blue), $\unit[12]{MeV}$ (green), and $\unit[14]{MeV}$ (red), compared to a composite combined radiograph (solid purple), modelled using a thermal proton distribution absorbed by a layer of RCF. c) The magnetic field profile reconstructed from the combined blurred radiograph (bold purple line) compared to the true field profile (solid black), and the reconstruction from the mono-energetic radiograph at $\unit[14]{MeV}$ (dashed red).}
\label{fig: Schematic}
\end{center}
\end{figure*}

We start by considering how protons of different energies passing through a significant background magnetic field are deflected by different angles depending on their energies, such as shown in Fig.~\ref{fig: Schematic}a). This results in a succession of different radiographs from protons of different energies, imprinted one on the other, each shifted by a small distance, as shown in Fig.~\ref{fig: Schematic}b). This plot shows proton radiographs calculated for the example magnetic field profile (with no electric fields) at three different distinct energies, from $\unit[10-14]{MeV}$, with each mono-energetic radiograph slightly displaced because of the background magnetic field. Each layer of RCF absorbs protons over a finite range of energies and so, when the proton beam has a broadband energy spectrum, the radiographs measured by each separate layer of film are all blurred in the direction of proton deflection.

In this example, the background field was chosen to resemble the conditions of a recent experiment. The peak field strength of $\unit[3]{T}$ along the $x$ direction extends over a diameter of around $\unit[40]{mm}$, as shown in Fig.~\ref{fig: Schematic}a), with the proton beam initially travelling along the $z$ axis and experiencing a total integrated field of $\int B_x \mathrm{d}z \approx \unit[90]{T mm}$; the proton beam is therefore deflected in the $y$ direction. The synthetic radiograph intensities were calculated using the EPOCH Particle-in-Cell code \cite{arber2015} and measured at an RCF position $\unit[100]{mm}$ from the origin. The radiographs were combined, weighted by a thermal proton spectrum at $T = \unit[5]{MeV}$ and by the calculated absorption of an RCF layer with an absorption peak at $\unit[10.6]{MeV}$ and a full-width-at-half-maximum of $\unit[0.4]{MeV}$. Note that this blurring is asymmetric and changes both the shape and the symmetry properties of the final radiograph. Using this combined radiograph to reconstruct the magnetic field profile therefore gives a poor estimate, as shown in Fig.~\ref{fig: Schematic}c). The amplitude of the signal reconstructed from the combined radiograph is much lower than the true value, as though the radiograph was taken with very poor spatial resolution. The shape of the profile is also distinctly different due to the asymmetric blurring. The reconstruction using a mono-energetic radiograph, on the other hand, is close to the true field, showing that the discrepancy arises from the blurring rather than from the reconstruction.

\begin{figure*}
\begin{center}
\includegraphics[scale=0.8]{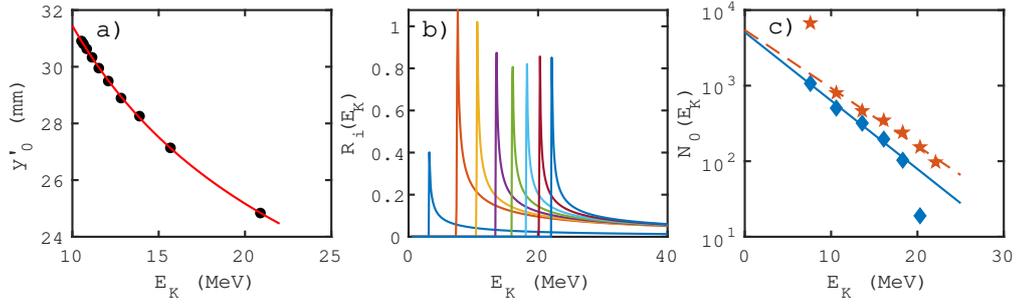}
\caption{a) Deflection of the proton beam against energy, measured by tracking the edge of the proton distribution in synthetic radiographs, and fitted to $y'_0(E) = a + b/\sqrt{E}$. b) RCF absorption curves calculated by the finite element opacity model described, with the first layer of RCF on the left. c) Example proton energy distributions estimated from RCF doses measured on two different experimental laser shots, fitted to thermal distributions.}
\label{fig: Energy Curves}
\end{center}
\end{figure*}

In order to calculate the degree of blurring in these synthetic radiographs we can consider each component in turn:

\emph{I) Proton Deflection}

First, we can calculate the deflection of protons in a known background magnetic field by tracking the path of the proton through the field, as shown in Fig.~\ref{fig: Schematic}a). On experiment, however, if the background field is not well characterised, the deflection must normally be calibrated using a known fiducial. By placing a wire at the interaction point and taking measurements of the position of the wire's shadow on different layers of RCF, we can estimate the vertical displacement of the beam on the stack $y'_0(E)$ as shown in Fig.~\ref{fig: Energy Curves}a). Strictly, this method assumes that the layers of RCF all lie in exactly the same place, as $y'_0$ also depends on the RCF position, but this assumption is accurate when the deflection angle $\theta_y$ is sufficiently small or the depth of the stack $\Delta z$ is sufficiently short that $\theta_y \Delta z \ll \Delta y'_0$. The measured points can then be fit to a function of the form $y'_0(E) \approx a + b/\sqrt{E}$, with fitted parameters $a$ and $b$, to reflect the higher deflection for lower energy protons. The steeper the change in $y'_0(E)$ with changes in proton energy, the worse the effect of blurring will be.

\emph{II) Protons Absorbed}

Secondly, the energies deposited into the RCF stack by the proton beam can be calculated using a finite element opacity model, where for a number of protons $N(E,z)$ of a given energy $E$ and depth in the stack $z$, the absorption in the stack is given by $\mathrm{d}N(E,z) = - \sigma(E,z) N(E,z) \mathrm{d}z$ for an opacity $\sigma(E,z)$. $\sigma$ can be tabulated against proton energy for different materials, with opacity higher for denser materials and for lower proton energies. The proton populations through the RCF stack can be calculated by integration as $\log \left [ N(E,z) / N_0(E) \right ] = - \int \sigma(E,z) \mathrm{d}z$, for an initial proton energy spectrum $N_0(E)$. The response function for a given slice of RCF labeled $i$ with thickness $t_i$ is then $R_i(E)=\left[N(E,z_i)-N(E,z_i+t_i)\right]/N_0(E)$. These response functions are plotted in Fig.~\ref{fig: Energy Curves}b), showing the characteristic Bragg peaks. For each layer of RCF this results in a negligible response at low energies (where almost all the protons have already been deposited earlier in the stack), followed by a sudden spike in absorption at a particular energy (where the integrated opacity is close to 1), followed by a decay in response at higher energies (where the integrated opacity remains much less than 1).

\emph{III) Proton Spectrum}

Finally, the response curves, which vary for each layer in the RCF stack, should then be weighted by the energy distribution of the proton beam $N_0(E)$, which is the same for all layers but varies shot to shot. The spectrum is generally estimated using the dose measured on each piece of RCF, compared with the proton energy most strongly absorbed by each layer. Either fitting to an expected spectral shape (such as a thermal spectrum given by $N_0(E) \propto e^{-E/T}$ for a temperature $T$) or interpolating between these measured points on the spectrum then gives an estimate of the proton energy distribution, with examples from two different laser shots shown in Fig.~\ref{fig: Energy Curves}c) with estimated temperatures of $\unit[4.8]{MeV}$ and $\unit[5.7]{MeV}$. Strictly, however, each layer of RCF absorbs protons from a range of energies, as we saw, and the proton spectrum should be found self-consistently, finding a spectrum $N_0(E)$ such that the measured dose on each slice $i$ is $D_i=\int R_i(E)N_0(E) \mathrm{dE}$. This can be approached by inversion or iteratively, starting at the back of the RCF stack where the response is only due to high energies and $R_i(E)$ is mostly zero.

\section{Blurring as a convolution}


Now we understand the causes of blurring in proton radiographs taken through a strong background magnetic field, we can work to remove the effect and recover an estimate of what the radiographs would look like without the background field. If the proton energy is conserved, in the absence of electric fields, the deflection of protons depends on the path-integrated magnetic field as $\mathbf{\hat{v_f}}-\mathbf{\hat{v_0}} = (q/\gamma m v) \int \mathbf{\hat{v}} \times \mathbf{B} ~\mathrm{d}s$. If the spatial scale of the background field is much larger than that of the field we wish to measure, the path of the protons is dominated by the applied field and we can safely calculate the degree of blurring from the background field alone.

On the other hand, the intensity of the proton radiograph is dependent on the gradient of the deflection. A one dimensional radiograph has an intensity profile given by $I/I_0 = |\partial y' / \partial y|^{-1}$, where $y$ is the position of protons as they pass through the object plane at $z=0$, and $y'$ is the position of protons as they arrive at the RCF stack. This means that the relevant quantity for the intensity profile is $(q/\gamma m v) \partial (\int \mathbf{\hat{v}} \times \mathbf{B} ~\mathrm{d}s) / \partial y$. If the background field is only slowly varying in space, the intensity profile of the radiograph is dominated by the lower amplitude but faster varying field we wish to measure (the signal). The shape of an un-blurred mono-energetic radiograph then only depends on the signal magnetic field.

Next, we assume that over the energy range of protons absorbed by a single layer of RCF, the radiograph is identical. Generally, lower energy protons are deflected more by the magnetic fields, giving radiographs with higher variations in intensity and more caustic features. At low proton energies or strong fields the assumption of identical radiographs over a small energy range is therefore not accurate. Where each RCF layer is only absorbing protons from a relatively narrow energy spread, however, it means we can write the intensity profile from protons of a given energy on a given shot as:
\begin{align}
 I_j(y';E) = I_{j0} \exp\left(-\frac{E}{T_j}\right) f\left[ y' - y'_0(E); E_0 \right],
\end{align}
where the proton spectrum on a laser shot~$j$ is characterised by an intensity $I_{j,0}$ and a temperature $T_j$ which vary shot-to-shot; the function $f(y';E_0)$ describes the shape of a mono-energetic radiograph at an energy $E_0 \approx E$, which depends only on the signal magnetic field; and $y'_0(E)$ describes the deflection which depends only on the background magnetic field.

When these protons are absorbed by a given RCF layer~$i$, they produce a measured dose shape of the form:
\begin{align}
 D_{ij}(y) &= I_{j,0} \int R_i(E) \exp\left(-\frac{E}{T_j}\right) f\left[ y' - y'_0(E) ;E_0 \right] \mathrm{d}E \\
        &= D_{j,0} \int g_{ij}(y'_0) f \left[ y' - y'_0 ;E_0 \right] \mathrm{d} y'_0
\end{align}
which is a linear spatial convolution of the mono-energetic radiograph with a kernel $g(y')$ given by:
\begin{align}
 g_{ij}(y') = R_i\left[ E(y') \right] \exp\left[-\frac{E(y')}{T_j}\right] \frac{\mathrm{d}E}{\mathrm{d}y'}
\end{align}

Furthermore, if the contrast of the radiograph is not so high as to form caustic features (i.e. $I/I_0 - 1 \ll 1$), we can generalise this convolution kernel to any RCF position or proton source location. If there is a direct mapping between points in the object plane described by the co-ordinate $y$ and points on the RCF stack described by $y' \approx y'_0(E) + My$ (for a magnification $M$) we can also describe the convolution kernel in terms of $y$. The width of the convolution kernel then describes the spatial resolution with which we can measure the magnetic field.

\begin{figure}
\begin{center}
\includegraphics[scale=0.9]{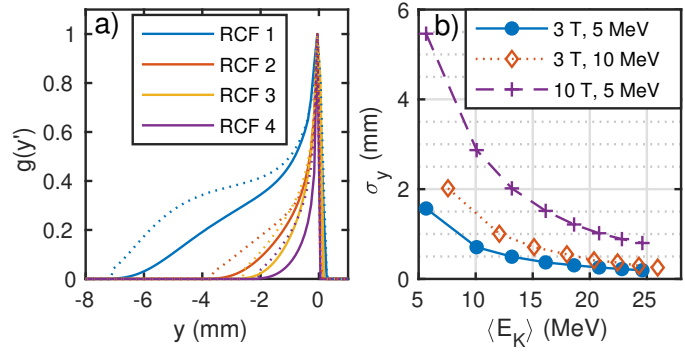}
\caption{a) Example convolution kernels calculated for different RCF layers for a $T=\unit[5]{MeV}$ thermal proton spectrum (solid lines) passing through a $\unit[3]{T}$ background field, compared to the kernels for a $T=\unit[10]{MeV}$ proton spectrum under the same conditions (dotted lines). b) The standard deviation of the kernels plotted against the mean proton energy absorbed by each RCF layer, comparing proton beams with $T=\unit[5]{MeV}$ (solid blue circles) and $T=\unit[10]{MeV}$ (dotted red diamonds) passing through a $\unit[3]{T}$ background field, to a $T=\unit[5]{MeV}$ proton beam passing through a $\unit[10]{T}$ background field. }
\label{fig: Kernel Comparison}
\end{center}
\end{figure}

Some example convolution kernels are shown in Fig.~\ref{fig: Kernel Comparison}a), calculated using the measured deflection and RCF response functions shown in Fig.~\ref{fig: Energy Curves}, and thermal proton energy spectra at $T=\unit[5]{MeV}$ and $T=\unit[10]{MeV}$, and plotted against the position in the object plane. For the first RCF layer at the front of the stack, with proton energies around $\unit[5]{MeV}$, the proton deflection is large and the resulting blurring kernel has a broad tail to the left of the ideal mono-energetic peak. This would blur out any magnetic field features smaller than a few millimeters. By the fourth RCF layer in the stack, however, with proton energies around $\unit[15]{MeV}$, the effect of blurring occurs over less than a millimeter.

Comparing the results for a proton spectrum with $T=\unit[10]{MeV}$, the broader energy range of protons incident on the RCF stack here results in broader blurring kernels. There is therefore a trade-off with proton energy: for higher energy protons deflection is small and blurring is negligible, but the dose absorbed in the RCF is lower and the amplitude of the signal will also be smaller. Increasing the dose by achieving higher temperature proton spectra will also increase the effect of blurring.

We can compare the effective spatial resolutions by looking at the standard deviations of the convolution kernels, plotted in Fig.~\ref{fig: Kernel Comparison}b) against the mean proton energy absorbed by each layer of the RCF stack. This shows the reduction in the kernel width with proton energy, such that for a $\unit[3]{T}$ background field with a field integral of $\unit[90]{T mm}$, the blurring width changes from around $\unit[2]{mm}$ in the object plane at $\unit[5]{MeV}$ to under $\unit[250]{\mu m}$ above $\unit[20]{MeV}$. Increasing the temperature of the proton beam to $\unit[10]{MeV}$ could increase the proton dose at these higher energies, but also leads to an increase of the blurring width by around $50\%$. If the field strength is increased to $\unit[10]{T}$, on the other hand, the effect of blurring is very significant even at these higher energies, with a blurring width of $\unit[1]{mm}$ in the object plane at $\unit[20]{MeV}$. Whereas magnetised plasma experiments which employ proton radiography often attempt to increase the proton temperature or the magnetic field strength, both of these changes will lead to greater blurring, and great care must be taken if small features in the magnetic field are to be measured.

\section{Deconvolution}

\begin{figure}
\begin{center}
\includegraphics[scale=0.9,trim=0.5cm 0 0 0]{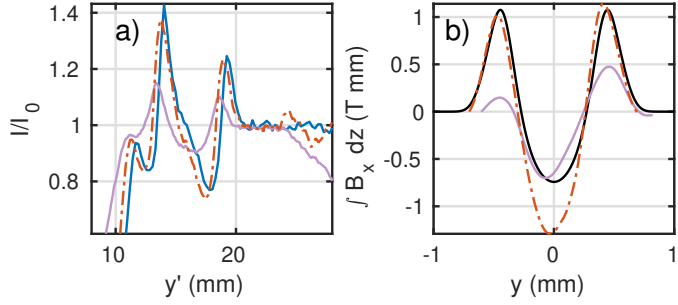}
\caption{a) Synthetic radiographs from a thermal proton beam passing through a background field with a strength of $\unit[90]{T mm}$ before being absorbed by a layer of RCF. The combined radiograph (faded purple line) and after applying a deconvolution (dash-dot red) are compared with a mono-energetic radiograph at a proton energy of $\unit[10]{MeV}$ (solid blue). b) The reconstructed magnetic field profile recovered from the synthetic radiographs, again comparing the result without the deconvolution (faded purple line) and the result using the deconvolution (dash-dot red), with the true magnetic field profile (solid black).}
\label{fig: Deconvolved}
\end{center}
\end{figure}

Having expressed the blurring as a linear convolution, it is possible to perform a deconvolution and recover the mono-energetic radiograph $f(y';E_0)$ from the measured dose profile. There are several possible deconvolution algorithms, of which we use the Richardson-Lucy technique \cite{richardson1972,lucy1974} for its stability. Fig.~\ref{fig: Deconvolved}a) shows the same combined radiograph described earlier, calculated from a thermal proton spectrum with $T=\unit[5]{MeV}$ incident on a layer of RCF after passing through a background magnetic field with a strength of $\unit[90]{T mm}$, as shown in Fig.~\ref{fig: Schematic}. This radiograph is compared with the $\unit[10]{MeV}$ mono-energetic radiograph predicted from particle-in-cell simulations, and the estimated radiograph recovered using deconvolution of the combined radiograph. This shows how the deconvolution increases the contrast of the radiograph, with the result closely approximating the profile of the mono-energetic radiograph. Under these conditions, this therefore makes the recovered magnetic field profile more accurate after using a deconvolution, as shown in Fig.~\ref{fig: Deconvolved}b). The deconvolved radiograph gives an estimate for the magnetic field which is almost an exact match to the true profile for $|r|>\unit[0.3]{mm}$, with the same symmetry and amplitude. On-axis, however, the deconvolution leads to a exaggeration of the magnetic field strength and a significant error, demonstrating that this deconvolution is a useful tool, but not a perfect solution to the problem of blurring.

Deconvolution of a blurred radiograph clearly has limitations, and cannot perfectly recover a mono-energetic radiograph or the true magnetic field profile. Firstly, applying any deconvolution algorithm to real data can amplify noise features, such as seen on the right of Fig.~\ref{fig: Deconvolved}a); instead of a flat $I=I_0$ profile there is an artefact from applying a deconvolution to noise. This can lead to inaccurate field reconstructions away from the main features, or where the amplitude of the noise is similar to that of the signal. Secondly, the shape of the radiograph is not identical for different proton energies, with the greater deflection of lower energy protons leading to a higher contrast radiograph. By assuming that the radiograph shape is constant, the deconvolved signal will tend to overestimate the amplitude of the magnetic field profile. Finally, the deconvolution process is only an estimate of the true deconvolution, and cannot perfectly recover the original signal. The broader the convolution kernel relative to the feature size of the signal, the more poorly the deconvolution algorithm will perform. All of these limitations mean that a deconvolution is more helpful in estimating the true magnetic field profile in some conditions than in others.

\begin{figure*}
\begin{center}
\includegraphics[scale=0.9]{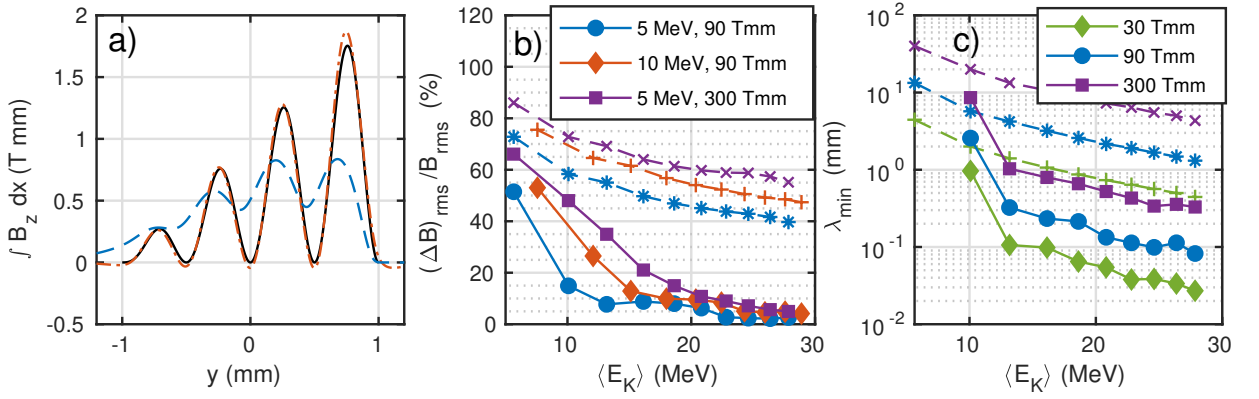}
\caption{a) An arbitrary integrated magnetic field profile with a scale length of $\unit[0.5]{mm}$ (solid black line) compared to an example reconstructed field from the blurred radiograph (dashed blue), and after employing the deconvolution (dash-dot red). b) The accuracy of the reconstruction before (dashed lines) and after (solid lines) the deconvolution, plotted against the mean proton energy absorbed in each layer of RCF, shown for thermal $T=\unit[5]{MeV}$ and $T=\unit[10]{MeV}$ protons passing through a background field of $\unit[90]{Tmm}$ or $\unit[300]{Tmm}$. c) An estimate of the maximum feature size that can be resolved to better than $10\%$ accuracy in the magnetic field profile with (solid lines, solid markers) and without the deconvolution (dashed lines, open markers), plotted against the mean proton energy absorbed in each layer of RCF, shown for a thermal $T=\unit[5]{MeV}$ proton spectrum passing through background fields of different strengths.}
\label{fig: Accuracy}
\end{center}
\end{figure*}

We can study the accuracy of the deconvolution under different conditions by simulating the blurring process and the deconvolution. Fig.~\ref{fig: Accuracy}a) shows an example of an arbitrary integrated magnetic field profile (solid black line), defined here as a sinusoid with a linearly increasing amplitude. The blurring is modelled by calculating a series of synthetic radiographs with different proton energies, before displacing the radiographs by the deflections shown in Fig.~\ref{fig: Energy Curves}a). This assumes that the background field is constant over the signal region and does not change the shape of the radiograph. The proton source is point-like in these ideal synthetic radiographs, with the spatial resolution limited only by the grid size of $\unit[2]{\mu m}$. The absorption of protons in the third layer of RCF is then modelled using a thermal proton spectrum with a temperature of $\unit[5]{MeV}$ and an RCF response curve as shown in Fig.~\ref{fig: Energy Curves}b). This results in significant blurring to the recovered magnetic field profile (dashed blue line), with the amplitude of the sinusoid reduced by around a factor of three. Applying a deconvolution algorithm, however, with a convolution kernel such as shown in Fig.~\ref{fig: Kernel Comparison}a), recovers a sharp radiograph. This gives an estimate of the magnetic field which closely matches the original profile.

Fig.~\ref{fig: Accuracy}b) shows how under different conditions the accuracy of the magnetic field reconstruction varies depending on the RCF layer, with layers which absorb higher energy protons measuring less blurring and giving a lower error. The error is shown as the root-mean-square difference in integrated magnetic field, relative to the root-mean-square of the original integrated magnetic field profile. Whereas the blurred radiographs (dashed lines) give relative errors of around $50\%$ or more, the error after applying the deconvolution (shown by the solid lines) falls to under $10\%$. The deconvolution is not perfect, however, and for the first layer in the RCF stack, absorbing proton energies around $\unit[5]{MeV}$, the relative error after the deconvolution is still around $50\%$. At these low proton energies, the large widths of the kernels make the deconvolution inaccurate, while the shape of the radiograph also changes rapidly with proton energy and the blurring is not well described by a convolution. We can therefore establish firstly that applying a deconvolution with the relevant kernel substantially improves the accuracy of reconstructing the magnetic field, and secondly that this deconvolution performs better at higher proton energies, above $\unit[10]{MeV}$.

By changing the modelled deflection and proton spectrum we can also explore how the strength of the background field and temperature of the proton beam affect the accuracy of the reconstruction. As expected, and in agreement with Fig.~\ref{fig: Kernel Comparison}b), both a higher background field and a higher temperature proton beam cause greater blurring and a reduction in accuracy of the reconstruction, both before and after applying a deconvolution. Whereas increasing the proton beam temperature from $\unit[5]{MeV}$ (shown by the blue lines) to $\unit[10]{MeV}$ (the red lines) makes the relative error without any deconvolution around $10\%$ worse, the main effect on the deconvolution process is to increase the proton energy required for an accurate reconstruction. A higher proton temperature not only increases the width of the smearing kernel, but also weakens the assumptions of a linear convolution. The shape of the radiograph can vary considerably over the larger energy range, making the deconvolution inaccurate below $\unit[15]{MeV}$. Above this point, however, the assumptions hold and the kernel is narrower, making the deconvolution accurate to better than $10\%$.

Similarly, increasing the background field strength from $\unit[90]{T mm}$ to $\unit[300]{T mm}$ (shown by the purple lines) makes the convolution kernels much broader, and increases the relative error on the blurred reconstruction by around $20\%$. The kernels for RCF layers at the front of the stack have widths greater than $\unit[2]{mm}$ in the object plane, as shown in Fig.~\ref{fig: Kernel Comparison}b), and it is therefore unsurprising that the deconvolution cannot accurately recover the features of the true magnetic field profile. By RCF layers deeper in the stack absorbing higher proton energies, on the other hand, the deconvolution accurately reproduces the true magnetic field profile, with a relative error of less than $10\%$ for proton energies greater than $\unit[20]{MeV}$. In this way, accurately probing magnetic fields in the presence of a strong background field requires high doses of high energy protons, but unfortunately using high temperature proton beams to achieve this can be counter-productive. Deconvolution is very successful at recovering the signal magnetic field profile even when blurring is severe, but spatial features that are too small are lost -- a feature we explore in detail next.

We can estimate the spatial resolution of the deconvolved radiographs by calculating the radiograph and the reconstructed magnetic field profile as before, now using a series of magnetic field profiles with the same form as shown in Fig.~\ref{fig: Accuracy}a) but with wavelengths from $\unit[16]{mm}$ down to $\unit[12.5]{\mu m}$. For each layer of RCF, the smallest wavelength is found which still gives less than $10\%$ root-mean-square error. The accuracy of the blurred radiographs, on the other hand, is estimated using the convolution kernel by finding the wavenumber where the Fourier transform of the relevant kernel falls to $90\%$ of the maximum amplitude, therefore introducing a $10\%$ error on the corresponding wavelength. 

The resulting estimates of the spatial resolution in the object plane are shown in Fig.~\ref{fig: Accuracy}c), with the maximum spatial resolution that can be recovered with an accuracy of better than $10\%$ plotted against proton energy for synthetic radiographs, both before and after applying the deconvolution, and for different background magnetic field strengths. This shows how the blurring strongly limits the spatial resolution, with this effect worse at lower proton energies, but still problematic for proton energies above $\unit[20]{MeV}$. At $\unit[20]{MeV}$, the spatial resolution is limited to around a millimeter for a background field strength of $\unit[30]{Tmm}$, but around $10$ millimeters for a background field strength of $\unit[300]{Tmm}$.

Applying the deconvolution improves the spatial resolution above $\unit[10]{MeV}$ by around an order of magnitude. For a $\unit[30]{Tmm}$ background field, spatial resolutions of 10s of microns in the object plane are achievable for proton energies higher than $\unit[20]{MeV}$. The deconvolution is not perfect and cannot recover all of the lost resolution, with the front layers of the RCF stack experiencing little benefit when absorbing protons around $\unit[10]{MeV}$ and below. Higher proton energies which experience less deflection again correspond to better spatial resolution. For the strongest $\unit[300]{Tmm}$ background field the spatial resolution even after using the deconvolution is around a millimeter at $\unit[15]{MeV}$, or $\unit[300]{\mu m}$ at $\unit[25]{MeV}$. When using a proton source with a broad energy spread to conduct radiography in applied magnetic fields, increasing the magnetic field strength will reduce the spatial resolution achievable, even after using a deconvolution.

\section{Conclusions}


We have extended proton probing to a new class of experiments which use applied magnetic fields surrounding the region of interest. These experiments encounter significant difficulties, because not only is the proton beam deflected by the background field, but this deflection is energy-dependent. When combined with a broadband proton energy spectrum and absorption in layers of radiochromic film, this deflection results in significant blurring of the proton radiograph. This blurring is most severe for layers of RCF at the front of the stack, which absorb lower energy protons, but is also a significant problem for protons of higher energies, above $\unit[20]{MeV}$ for the conditions considered here. Furthermore, increasing the temperature of the proton beam in order to access higher proton energies will itself worsen the effect of blurring.

However, we have also shown that under certain conditions the blurring can be modelled as a linear convolution and removed using a deconvolution algorithm. When the background field is large and slowly varying in space compared with the signal, and the relative energy spread of protons absorbed by the RCF is sufficiently small, we can therefore recover a good estimate for a mono-energetic radiograph and accurately reconstruct the magnetic field we wish to measure despite the presence of a strong background field. By looking at how the convolution kernel changes with the background field strength, the proton temperature, and the RCF absorption, we can estimate the loss of spatial resolution caused by the blurring and calculate the deconvolution required.

We have showed how a deconvolution substantially increases the accuracy of the reconstructed magnetic field profile, with the error on the reconstruction falling from over $50\%$ to under $10\%$, and the spatial resolution of the radiographs is improved by an order of magnitude. While no estimate for the true magnetic field profile will be perfect, calculating the kernel and applying the deconvolution allows us to recover much greater spatial resolution than possible from the blurred radiographs, down to around $\unit[100]{\mu m}$ for a $\unit[90]{Tmm}$ background field. We have in this way extended the power of proton radiography to experiments with applied magnetic fields, allowing researchers to study changes to the electric and magnetic fields even under these challenging experimental conditions.

\begin{acknowledgments}
All authors contributed to this work.

The authors are grateful for the support of LLNL Academic Partnerships (B618488), EUROfusion Enabling Research Grants  AWP17-ENR-IFE-CCFE-01 and AWP17-ENR-IFE-CEA-02, and UK EPSRC grants EP/P026796/1 and EP/R029148/1.

The code and data used in this study are available from the corresponding author upon reasonable request.
\end{acknowledgments}


\begin{thebibliography}{21}%
\makeatletter
\providecommand \@ifxundefined [1]{%
 \@ifx{#1\undefined}
}%
\providecommand \@ifnum [1]{%
 \ifnum #1\expandafter \@firstoftwo
 \else \expandafter \@secondoftwo
 \fi
}%
\providecommand \@ifx [1]{%
 \ifx #1\expandafter \@firstoftwo
 \else \expandafter \@secondoftwo
 \fi
}%
\providecommand \natexlab [1]{#1}%
\providecommand \enquote  [1]{``#1''}%
\providecommand \bibnamefont  [1]{#1}%
\providecommand \bibfnamefont [1]{#1}%
\providecommand \citenamefont [1]{#1}%
\providecommand \href@noop [0]{\@secondoftwo}%
\providecommand \href [0]{\begingroup \@sanitize@url \@href}%
\providecommand \@href[1]{\@@startlink{#1}\@@href}%
\providecommand \@@href[1]{\endgroup#1\@@endlink}%
\providecommand \@sanitize@url [0]{\catcode `\\12\catcode `\$12\catcode
  `\&12\catcode `\#12\catcode `\^12\catcode `\_12\catcode `\%12\relax}%
\providecommand \@@startlink[1]{}%
\providecommand \@@endlink[0]{}%
\providecommand \url  [0]{\begingroup\@sanitize@url \@url }%
\providecommand \@url [1]{\endgroup\@href {#1}{\urlprefix }}%
\providecommand \urlprefix  [0]{URL }%
\providecommand \Eprint [0]{\href }%
\providecommand \doibase [0]{http://dx.doi.org/}%
\providecommand \bibinfo  [0]{\@secondoftwo}%
\providecommand \bibfield  [0]{\@secondoftwo}%
\providecommand \translation [1]{[#1]}%
\providecommand \BibitemOpen [0]{}%
\providecommand \bibitemStop [0]{}%
\providecommand \bibitemNoStop [0]{.\EOS\space}%
\providecommand \EOS [0]{\spacefactor3000\relax}%
\providecommand \BibitemShut  [1]{\csname bibitem#1\endcsname}%
\let\auto@bib@innerbib\@empty
\bibitem [{\citenamefont {Borghesi}\ \emph {et~al.}(2006)\citenamefont
  {Borghesi}, \citenamefont {Fuchs}, \citenamefont {Bulanov}, \citenamefont
  {MacKinnon}, \citenamefont {Patel},\ and\ \citenamefont
  {Roth}}]{borghesi2006}%
  \BibitemOpen
  \bibfield  {author} {\bibinfo {author} {\bibfnamefont {M.}~\bibnamefont
  {Borghesi}}, \bibinfo {author} {\bibfnamefont {J.}~\bibnamefont {Fuchs}},
  \bibinfo {author} {\bibfnamefont {S.~V.}\ \bibnamefont {Bulanov}}, \bibinfo
  {author} {\bibfnamefont {A.~J.}\ \bibnamefont {MacKinnon}}, \bibinfo {author}
  {\bibfnamefont {P.~K.}\ \bibnamefont {Patel}}, \ and\ \bibinfo {author}
  {\bibfnamefont {M.}~\bibnamefont {Roth}},\ }\bibfield  {title}
  {{ \enquote {\bibinfo {title} {Fast {{Ion Generation}} by
  {{High}}-{{Intensity Laser Irradiation}} of {{Solid Targets}} and
  {{Applications}}},}\ }}\href {\doibase 10.13182/FST06-A1159} {\bibfield
  {journal} {\bibinfo  {journal} {Fusion Science and Technology}\ }\textbf
  {\bibinfo {volume} {49}},\ \bibinfo {pages} {412--439} (\bibinfo {year}
  {2006})}\BibitemShut {NoStop}%
\bibitem [{\citenamefont {Kugland}\ \emph {et~al.}(2012)\citenamefont
  {Kugland}, \citenamefont {Ryutov}, \citenamefont {Plechaty}, \citenamefont
  {Ross},\ and\ \citenamefont {Park}}]{kugland2012}%
  \BibitemOpen
  \bibfield  {author} {\bibinfo {author} {\bibfnamefont {N.~L.}\ \bibnamefont
  {Kugland}}, \bibinfo {author} {\bibfnamefont {D.~D.}\ \bibnamefont {Ryutov}},
  \bibinfo {author} {\bibfnamefont {C.}~\bibnamefont {Plechaty}}, \bibinfo
  {author} {\bibfnamefont {J.~S.}\ \bibnamefont {Ross}}, \ and\ \bibinfo
  {author} {\bibfnamefont {H.-S.}\ \bibnamefont {Park}},\ }\bibfield  {title}
  {\enquote {\bibinfo {title} {Invited {{Article}}: {{Relation}} between
  electric and magnetic field structures and their proton-beam images},}\
  }\href {\doibase 10.1063/1.4750234} {\bibfield  {journal} {\bibinfo
  {journal} {Review of Scientific Instruments}\ }\textbf {\bibinfo {volume}
  {83}},\ \bibinfo {pages} {101301} (\bibinfo {year} {2012})}\BibitemShut
  {NoStop}%
\bibitem [{\citenamefont {Kasim}\ \emph {et~al.}(2019)\citenamefont {Kasim},
  \citenamefont {Bott}, \citenamefont {Tzeferacos}, \citenamefont {Lamb},
  \citenamefont {Gregori},\ and\ \citenamefont {Vinko}}]{kasim2019}%
  \BibitemOpen
  \bibfield  {author} {\bibinfo {author} {\bibfnamefont {M.~F.}\ \bibnamefont
  {Kasim}}, \bibinfo {author} {\bibfnamefont {A.~F.~A.}\ \bibnamefont {Bott}},
  \bibinfo {author} {\bibfnamefont {P.}~\bibnamefont {Tzeferacos}}, \bibinfo
  {author} {\bibfnamefont {D.~Q.}\ \bibnamefont {Lamb}}, \bibinfo {author}
  {\bibfnamefont {G.}~\bibnamefont {Gregori}}, \ and\ \bibinfo {author}
  {\bibfnamefont {S.~M.}\ \bibnamefont {Vinko}},\ }\bibfield  {title} {\enquote
  {\bibinfo {title} {Retrieving fields from proton radiography without source
  profiles},}\ }\href {\doibase 10.1103/PhysRevE.100.033208} {\bibfield
  {journal} {\bibinfo  {journal} {Phys. Rev. E}\ }\textbf {\bibinfo {volume}
  {100}},\ \bibinfo {pages} {033208} (\bibinfo {year} {2019})}\BibitemShut
  {NoStop}%
\bibitem [{\citenamefont {Willingale}\ \emph {et~al.}(2010)\citenamefont
  {Willingale}, \citenamefont {Thomas}, \citenamefont {Nilson}, \citenamefont
  {Kaluza}, \citenamefont {Bandyopadhyay}, \citenamefont {Dangor},
  \citenamefont {Evans}, \citenamefont {Fernandes}, \citenamefont {Haines},
  \citenamefont {Kamperidis}, \citenamefont {Kingham}, \citenamefont {Minardi},
  \citenamefont {Notley}, \citenamefont {Ridgers}, \citenamefont {Rozmus},
  \citenamefont {Sherlock}, \citenamefont {Tatarakis}, \citenamefont {Wei},
  \citenamefont {Najmudin},\ and\ \citenamefont
  {Krushelnick}}]{willingale2010}%
  \BibitemOpen
  \bibfield  {author} {\bibinfo {author} {\bibfnamefont {L.}~\bibnamefont
  {Willingale}}, \bibinfo {author} {\bibfnamefont {A.~G.~R.}\ \bibnamefont
  {Thomas}}, \bibinfo {author} {\bibfnamefont {P.~M.}\ \bibnamefont {Nilson}},
  \bibinfo {author} {\bibfnamefont {M.~C.}\ \bibnamefont {Kaluza}}, \bibinfo
  {author} {\bibfnamefont {S.}~\bibnamefont {Bandyopadhyay}}, \bibinfo {author}
  {\bibfnamefont {A.~E.}\ \bibnamefont {Dangor}}, \bibinfo {author}
  {\bibfnamefont {R.~G.}\ \bibnamefont {Evans}}, \bibinfo {author}
  {\bibfnamefont {P.}~\bibnamefont {Fernandes}}, \bibinfo {author}
  {\bibfnamefont {M.~G.}\ \bibnamefont {Haines}}, \bibinfo {author}
  {\bibfnamefont {C.}~\bibnamefont {Kamperidis}}, \bibinfo {author}
  {\bibfnamefont {R.~J.}\ \bibnamefont {Kingham}}, \bibinfo {author}
  {\bibfnamefont {S.}~\bibnamefont {Minardi}}, \bibinfo {author} {\bibfnamefont
  {M.}~\bibnamefont {Notley}}, \bibinfo {author} {\bibfnamefont {C.~P.}\
  \bibnamefont {Ridgers}}, \bibinfo {author} {\bibfnamefont {W.}~\bibnamefont
  {Rozmus}}, \bibinfo {author} {\bibfnamefont {M.}~\bibnamefont {Sherlock}},
  \bibinfo {author} {\bibfnamefont {M.}~\bibnamefont {Tatarakis}}, \bibinfo
  {author} {\bibfnamefont {M.~S.}\ \bibnamefont {Wei}}, \bibinfo {author}
  {\bibfnamefont {Z.}~\bibnamefont {Najmudin}}, \ and\ \bibinfo {author}
  {\bibfnamefont {K.}~\bibnamefont {Krushelnick}},\ }\bibfield  {title}
  {{ \enquote {\bibinfo {title} {Fast {{Advection}} of
  {{Magnetic Fields}} by {{Hot Electrons}}},}\ }}\href {\doibase
  10.1103/PhysRevLett.105.095001} {\bibfield  {journal} {\bibinfo  {journal}
  {Phys. Rev. Lett.}\ }\textbf {\bibinfo {volume} {105}},\ \bibinfo {pages}
  {095001} (\bibinfo {year} {2010})}\BibitemShut {NoStop}%
\bibitem [{\citenamefont {Li}\ \emph {et~al.}(2013)\citenamefont {Li},
  \citenamefont {S{\'e}guin}, \citenamefont {Frenje}, \citenamefont {Sinenian},
  \citenamefont {Rosenberg}, \citenamefont {Manuel}, \citenamefont
  {Rinderknecht}, \citenamefont {Zylstra}, \citenamefont {Petrasso},
  \citenamefont {Amendt}, \citenamefont {Landen}, \citenamefont {Mackinnon},
  \citenamefont {Town}, \citenamefont {Wilks}, \citenamefont {Betti},
  \citenamefont {Meyerhofer}, \citenamefont {Soures}, \citenamefont {Hund},
  \citenamefont {Kilkenny},\ and\ \citenamefont {Nikroo}}]{li2013}%
  \BibitemOpen
  \bibfield  {author} {\bibinfo {author} {\bibfnamefont {C.}~\bibnamefont
  {Li}}, \bibinfo {author} {\bibfnamefont {F.}~\bibnamefont {S{\'e}guin}},
  \bibinfo {author} {\bibfnamefont {J.}~\bibnamefont {Frenje}}, \bibinfo
  {author} {\bibfnamefont {N.}~\bibnamefont {Sinenian}}, \bibinfo {author}
  {\bibfnamefont {M.}~\bibnamefont {Rosenberg}}, \bibinfo {author}
  {\bibfnamefont {M.-E.}\ \bibnamefont {Manuel}}, \bibinfo {author}
  {\bibfnamefont {H.}~\bibnamefont {Rinderknecht}}, \bibinfo {author}
  {\bibfnamefont {A.}~\bibnamefont {Zylstra}}, \bibinfo {author} {\bibfnamefont
  {R.}~\bibnamefont {Petrasso}}, \bibinfo {author} {\bibfnamefont
  {P.}~\bibnamefont {Amendt}}, \bibinfo {author} {\bibfnamefont
  {O.}~\bibnamefont {Landen}}, \bibinfo {author} {\bibfnamefont
  {A.}~\bibnamefont {Mackinnon}}, \bibinfo {author} {\bibfnamefont
  {R.}~\bibnamefont {Town}}, \bibinfo {author} {\bibfnamefont {S.}~\bibnamefont
  {Wilks}}, \bibinfo {author} {\bibfnamefont {R.}~\bibnamefont {Betti}},
  \bibinfo {author} {\bibfnamefont {D.}~\bibnamefont {Meyerhofer}}, \bibinfo
  {author} {\bibfnamefont {J.}~\bibnamefont {Soures}}, \bibinfo {author}
  {\bibfnamefont {J.}~\bibnamefont {Hund}}, \bibinfo {author} {\bibfnamefont
  {J.}~\bibnamefont {Kilkenny}}, \ and\ \bibinfo {author} {\bibfnamefont
  {A.}~\bibnamefont {Nikroo}},\ }\bibfield  {title} {{ \enquote {\bibinfo {title} {Proton imaging of hohlraum plasma stagnation
  in inertial-confinement-fusion experiments},}\ }}\href {\doibase
  10.1088/0029-5515/53/7/073022} {\bibfield  {journal} {\bibinfo  {journal}
  {Nucl. Fusion}\ }\textbf {\bibinfo {volume} {53}},\ \bibinfo {pages} {073022}
  (\bibinfo {year} {2013})}\BibitemShut {NoStop}%
\bibitem [{\citenamefont {Palmer}\ \emph {et~al.}(2019)\citenamefont {Palmer},
  \citenamefont {Campbell}, \citenamefont {Ma}, \citenamefont {Antonelli},
  \citenamefont {Bott}, \citenamefont {Gregori}, \citenamefont {Halliday},
  \citenamefont {Katzir}, \citenamefont {Kordell}, \citenamefont {Krushelnick},
  \citenamefont {Lebedev}, \citenamefont {Montgomery}, \citenamefont {Notley},
  \citenamefont {Carroll}, \citenamefont {Ridgers}, \citenamefont
  {Schekochihin}, \citenamefont {Streeter}, \citenamefont {Thomas},
  \citenamefont {Tubman}, \citenamefont {Woolsey},\ and\ \citenamefont
  {Willingale}}]{palmer2019}%
  \BibitemOpen
  \bibfield  {author} {\bibinfo {author} {\bibfnamefont {C.~a.~J.}\
  \bibnamefont {Palmer}}, \bibinfo {author} {\bibfnamefont {P.~T.}\
  \bibnamefont {Campbell}}, \bibinfo {author} {\bibfnamefont {Y.}~\bibnamefont
  {Ma}}, \bibinfo {author} {\bibfnamefont {L.}~\bibnamefont {Antonelli}},
  \bibinfo {author} {\bibfnamefont {A.~F.~A.}\ \bibnamefont {Bott}}, \bibinfo
  {author} {\bibfnamefont {G.}~\bibnamefont {Gregori}}, \bibinfo {author}
  {\bibfnamefont {J.}~\bibnamefont {Halliday}}, \bibinfo {author}
  {\bibfnamefont {Y.}~\bibnamefont {Katzir}}, \bibinfo {author} {\bibfnamefont
  {P.}~\bibnamefont {Kordell}}, \bibinfo {author} {\bibfnamefont
  {K.}~\bibnamefont {Krushelnick}}, \bibinfo {author} {\bibfnamefont {S.~V.}\
  \bibnamefont {Lebedev}}, \bibinfo {author} {\bibfnamefont {E.}~\bibnamefont
  {Montgomery}}, \bibinfo {author} {\bibfnamefont {M.}~\bibnamefont {Notley}},
  \bibinfo {author} {\bibfnamefont {D.~C.}\ \bibnamefont {Carroll}}, \bibinfo
  {author} {\bibfnamefont {C.~P.}\ \bibnamefont {Ridgers}}, \bibinfo {author}
  {\bibfnamefont {A.~A.}\ \bibnamefont {Schekochihin}}, \bibinfo {author}
  {\bibfnamefont {M.~J.~V.}\ \bibnamefont {Streeter}}, \bibinfo {author}
  {\bibfnamefont {A.~G.~R.}\ \bibnamefont {Thomas}}, \bibinfo {author}
  {\bibfnamefont {E.~R.}\ \bibnamefont {Tubman}}, \bibinfo {author}
  {\bibfnamefont {N.}~\bibnamefont {Woolsey}}, \ and\ \bibinfo {author}
  {\bibfnamefont {L.}~\bibnamefont {Willingale}},\ }\bibfield  {title}
  {\enquote {\bibinfo {title} {Field reconstruction from proton radiography of
  intense laser driven magnetic reconnection},}\ }\href {\doibase
  10.1063/1.5092733} {\bibfield  {journal} {\bibinfo  {journal} {Physics of
  Plasmas}\ }\textbf {\bibinfo {volume} {26}},\ \bibinfo {pages} {083109}
  (\bibinfo {year} {2019})}\BibitemShut {NoStop}%
\bibitem [{\citenamefont {Tubman}\ \emph {et~al.}(2021)\citenamefont {Tubman},
  \citenamefont {Joglekar}, \citenamefont {Bott}, \citenamefont {Borghesi},
  \citenamefont {Coleman}, \citenamefont {Cooper}, \citenamefont {Danson},
  \citenamefont {Durey}, \citenamefont {Foster}, \citenamefont {Graham},
  \citenamefont {Gregori}, \citenamefont {Gumbrell}, \citenamefont {Hill},
  \citenamefont {Hodge}, \citenamefont {Kar}, \citenamefont {Kingham},
  \citenamefont {Read}, \citenamefont {Ridgers}, \citenamefont {Skidmore},
  \citenamefont {Spindloe}, \citenamefont {Thomas}, \citenamefont {Treadwell},
  \citenamefont {Wilson}, \citenamefont {Willingale},\ and\ \citenamefont
  {Woolsey}}]{tubman2021}%
  \BibitemOpen
  \bibfield  {author} {\bibinfo {author} {\bibfnamefont {E.~R.}\ \bibnamefont
  {Tubman}}, \bibinfo {author} {\bibfnamefont {A.~S.}\ \bibnamefont
  {Joglekar}}, \bibinfo {author} {\bibfnamefont {A.~F.~A.}\ \bibnamefont
  {Bott}}, \bibinfo {author} {\bibfnamefont {M.}~\bibnamefont {Borghesi}},
  \bibinfo {author} {\bibfnamefont {B.}~\bibnamefont {Coleman}}, \bibinfo
  {author} {\bibfnamefont {G.}~\bibnamefont {Cooper}}, \bibinfo {author}
  {\bibfnamefont {C.~N.}\ \bibnamefont {Danson}}, \bibinfo {author}
  {\bibfnamefont {P.}~\bibnamefont {Durey}}, \bibinfo {author} {\bibfnamefont
  {J.~M.}\ \bibnamefont {Foster}}, \bibinfo {author} {\bibfnamefont
  {P.}~\bibnamefont {Graham}}, \bibinfo {author} {\bibfnamefont
  {G.}~\bibnamefont {Gregori}}, \bibinfo {author} {\bibfnamefont {E.~T.}\
  \bibnamefont {Gumbrell}}, \bibinfo {author} {\bibfnamefont {M.~P.}\
  \bibnamefont {Hill}}, \bibinfo {author} {\bibfnamefont {T.}~\bibnamefont
  {Hodge}}, \bibinfo {author} {\bibfnamefont {S.}~\bibnamefont {Kar}}, \bibinfo
  {author} {\bibfnamefont {R.~J.}\ \bibnamefont {Kingham}}, \bibinfo {author}
  {\bibfnamefont {M.}~\bibnamefont {Read}}, \bibinfo {author} {\bibfnamefont
  {C.~P.}\ \bibnamefont {Ridgers}}, \bibinfo {author} {\bibfnamefont
  {J.}~\bibnamefont {Skidmore}}, \bibinfo {author} {\bibfnamefont
  {C.}~\bibnamefont {Spindloe}}, \bibinfo {author} {\bibfnamefont {A.~G.~R.}\
  \bibnamefont {Thomas}}, \bibinfo {author} {\bibfnamefont {P.}~\bibnamefont
  {Treadwell}}, \bibinfo {author} {\bibfnamefont {S.}~\bibnamefont {Wilson}},
  \bibinfo {author} {\bibfnamefont {L.}~\bibnamefont {Willingale}}, \ and\
  \bibinfo {author} {\bibfnamefont {N.~C.}\ \bibnamefont {Woolsey}},\
  }\bibfield  {title} {{ \enquote {\bibinfo {title}
  {Observations of pressure anisotropy effects within semi-collisional
  magnetized plasma bubbles},}\ }}\href {\doibase 10.1038/s41467-020-20387-7}
  {\bibfield  {journal} {\bibinfo  {journal} {Nat. Commun.}\ }\textbf {\bibinfo
  {volume} {12}},\ \bibinfo {pages} {334} (\bibinfo {year} {2021})}\BibitemShut
  {NoStop}%
\bibitem [{\citenamefont {Gregori}\ \emph {et~al.}(2012)\citenamefont
  {Gregori}, \citenamefont {Ravasio}, \citenamefont {Murphy}, \citenamefont
  {Schaar}, \citenamefont {Baird}, \citenamefont {Bell}, \citenamefont
  {{Benuzzi-Mounaix}}, \citenamefont {Bingham}, \citenamefont {Constantin},
  \citenamefont {Drake}, \citenamefont {Edwards}, \citenamefont {Everson},
  \citenamefont {Gregory}, \citenamefont {Kuramitsu}, \citenamefont {Lau},
  \citenamefont {Mithen}, \citenamefont {Niemann}, \citenamefont {Park},
  \citenamefont {Remington}, \citenamefont {Reville}, \citenamefont {Robinson},
  \citenamefont {Ryutov}, \citenamefont {Sakawa}, \citenamefont {Yang},
  \citenamefont {Woolsey}, \citenamefont {Koenig},\ and\ \citenamefont
  {Miniati}}]{gregori2012}%
  \BibitemOpen
  \bibfield  {author} {\bibinfo {author} {\bibfnamefont {G.}~\bibnamefont
  {Gregori}}, \bibinfo {author} {\bibfnamefont {A.}~\bibnamefont {Ravasio}},
  \bibinfo {author} {\bibfnamefont {C.~D.}\ \bibnamefont {Murphy}}, \bibinfo
  {author} {\bibfnamefont {K.}~\bibnamefont {Schaar}}, \bibinfo {author}
  {\bibfnamefont {A.}~\bibnamefont {Baird}}, \bibinfo {author} {\bibfnamefont
  {A.~R.}\ \bibnamefont {Bell}}, \bibinfo {author} {\bibfnamefont
  {A.}~\bibnamefont {{Benuzzi-Mounaix}}}, \bibinfo {author} {\bibfnamefont
  {R.}~\bibnamefont {Bingham}}, \bibinfo {author} {\bibfnamefont
  {C.}~\bibnamefont {Constantin}}, \bibinfo {author} {\bibfnamefont {R.~P.}\
  \bibnamefont {Drake}}, \bibinfo {author} {\bibfnamefont {M.}~\bibnamefont
  {Edwards}}, \bibinfo {author} {\bibfnamefont {E.~T.}\ \bibnamefont
  {Everson}}, \bibinfo {author} {\bibfnamefont {C.~D.}\ \bibnamefont
  {Gregory}}, \bibinfo {author} {\bibfnamefont {Y.}~\bibnamefont {Kuramitsu}},
  \bibinfo {author} {\bibfnamefont {W.}~\bibnamefont {Lau}}, \bibinfo {author}
  {\bibfnamefont {J.}~\bibnamefont {Mithen}}, \bibinfo {author} {\bibfnamefont
  {C.}~\bibnamefont {Niemann}}, \bibinfo {author} {\bibfnamefont {H.-S.}\
  \bibnamefont {Park}}, \bibinfo {author} {\bibfnamefont {B.~A.}\ \bibnamefont
  {Remington}}, \bibinfo {author} {\bibfnamefont {B.}~\bibnamefont {Reville}},
  \bibinfo {author} {\bibfnamefont {A.~P.~L.}\ \bibnamefont {Robinson}},
  \bibinfo {author} {\bibfnamefont {D.~D.}\ \bibnamefont {Ryutov}}, \bibinfo
  {author} {\bibfnamefont {Y.}~\bibnamefont {Sakawa}}, \bibinfo {author}
  {\bibfnamefont {S.}~\bibnamefont {Yang}}, \bibinfo {author} {\bibfnamefont
  {N.~C.}\ \bibnamefont {Woolsey}}, \bibinfo {author} {\bibfnamefont
  {M.}~\bibnamefont {Koenig}}, \ and\ \bibinfo {author} {\bibfnamefont
  {F.}~\bibnamefont {Miniati}},\ }\bibfield  {title} { \enquote {\bibinfo {title} {Generation of scaled protogalactic seed
  magnetic fields in laser-produced shock waves},}\ }\href {\doibase
  10.1038/nature10747} {\bibfield  {journal} {\bibinfo  {journal} {Nature}\
  }\textbf {\bibinfo {volume} {481}},\ \bibinfo {pages} {480--483} (\bibinfo
  {year} {2012})}\BibitemShut {NoStop}%
\bibitem [{\citenamefont {Meinecke}\ \emph {et~al.}(2014)\citenamefont
  {Meinecke}, \citenamefont {Doyle}, \citenamefont {Miniati}, \citenamefont
  {Bell}, \citenamefont {Bingham}, \citenamefont {Crowston}, \citenamefont
  {Drake}, \citenamefont {Fatenejad}, \citenamefont {Koenig}, \citenamefont
  {Kuramitsu}, \citenamefont {C.~Kuranz}, \citenamefont {Lamb}, \citenamefont
  {Lee}, \citenamefont {MacDonald}, \citenamefont {Murphy}, \citenamefont
  {Park}, \citenamefont {Pelka}, \citenamefont {Ravasio}, \citenamefont
  {Sakawa}, \citenamefont {Schekochihin}, \citenamefont {Scopatz},
  \citenamefont {Tzeferacos}, \citenamefont {Wan}, \citenamefont {Woolsey},
  \citenamefont {Yurchak}, \citenamefont {Reville},\ and\ \citenamefont
  {Gregori}}]{meinecke2014}%
  \BibitemOpen
  \bibfield  {author} {\bibinfo {author} {\bibfnamefont {J.}~\bibnamefont
  {Meinecke}}, \bibinfo {author} {\bibfnamefont {H.~W.}\ \bibnamefont {Doyle}},
  \bibinfo {author} {\bibfnamefont {F.}~\bibnamefont {Miniati}}, \bibinfo
  {author} {\bibfnamefont {A.~R.}\ \bibnamefont {Bell}}, \bibinfo {author}
  {\bibfnamefont {R.}~\bibnamefont {Bingham}}, \bibinfo {author} {\bibfnamefont
  {R.}~\bibnamefont {Crowston}}, \bibinfo {author} {\bibfnamefont {R.~P.}\
  \bibnamefont {Drake}}, \bibinfo {author} {\bibfnamefont {M.}~\bibnamefont
  {Fatenejad}}, \bibinfo {author} {\bibfnamefont {M.}~\bibnamefont {Koenig}},
  \bibinfo {author} {\bibfnamefont {Y.}~\bibnamefont {Kuramitsu}}, \bibinfo
  {author} {\bibfnamefont {C.}~\bibnamefont {C.~Kuranz}}, \bibinfo {author}
  {\bibfnamefont {D.~Q.}\ \bibnamefont {Lamb}}, \bibinfo {author}
  {\bibfnamefont {D.}~\bibnamefont {Lee}}, \bibinfo {author} {\bibfnamefont
  {M.~J.}\ \bibnamefont {MacDonald}}, \bibinfo {author} {\bibfnamefont {C.~D.}\
  \bibnamefont {Murphy}}, \bibinfo {author} {\bibfnamefont {H.-S.}\
  \bibnamefont {Park}}, \bibinfo {author} {\bibfnamefont {A.}~\bibnamefont
  {Pelka}}, \bibinfo {author} {\bibfnamefont {A.}~\bibnamefont {Ravasio}},
  \bibinfo {author} {\bibfnamefont {Y.}~\bibnamefont {Sakawa}}, \bibinfo
  {author} {\bibfnamefont {A.~A.}\ \bibnamefont {Schekochihin}}, \bibinfo
  {author} {\bibfnamefont {A.}~\bibnamefont {Scopatz}}, \bibinfo {author}
  {\bibfnamefont {P.}~\bibnamefont {Tzeferacos}}, \bibinfo {author}
  {\bibfnamefont {W.~C.}\ \bibnamefont {Wan}}, \bibinfo {author} {\bibfnamefont
  {N.~C.}\ \bibnamefont {Woolsey}}, \bibinfo {author} {\bibfnamefont
  {R.}~\bibnamefont {Yurchak}}, \bibinfo {author} {\bibfnamefont
  {B.}~\bibnamefont {Reville}}, \ and\ \bibinfo {author} {\bibfnamefont
  {G.}~\bibnamefont {Gregori}},\ }\bibfield  {title} {{
  \enquote {\bibinfo {title} {Turbulent amplification of magnetic fields in
  laboratory laser-produced shock waves},}\ }}\href {\doibase
  10.1038/nphys2978} {\bibfield  {journal} {\bibinfo  {journal} {Nat. Phys.}\
  }\textbf {\bibinfo {volume} {10}},\ \bibinfo {pages} {520--524} (\bibinfo
  {year} {2014})}\BibitemShut {NoStop}%
\bibitem [{\citenamefont {Perkins}\ \emph {et~al.}(2017)\citenamefont
  {Perkins}, \citenamefont {Ho}, \citenamefont {Logan}, \citenamefont
  {Zimmerman}, \citenamefont {Rhodes}, \citenamefont {Strozzi}, \citenamefont
  {Blackfield},\ and\ \citenamefont {Hawkins}}]{perkins2017}%
  \BibitemOpen
  \bibfield  {author} {\bibinfo {author} {\bibfnamefont {L.~J.}\ \bibnamefont
  {Perkins}}, \bibinfo {author} {\bibfnamefont {D.~D.-M.}\ \bibnamefont {Ho}},
  \bibinfo {author} {\bibfnamefont {B.~G.}\ \bibnamefont {Logan}}, \bibinfo
  {author} {\bibfnamefont {G.~B.}\ \bibnamefont {Zimmerman}}, \bibinfo {author}
  {\bibfnamefont {M.~A.}\ \bibnamefont {Rhodes}}, \bibinfo {author}
  {\bibfnamefont {D.~J.}\ \bibnamefont {Strozzi}}, \bibinfo {author}
  {\bibfnamefont {D.~T.}\ \bibnamefont {Blackfield}}, \ and\ \bibinfo {author}
  {\bibfnamefont {S.~A.}\ \bibnamefont {Hawkins}},\ }\bibfield  {title}
  {\enquote {\bibinfo {title} {The potential of imposed magnetic fields for
  enhancing ignition probability and fusion energy yield in indirect-drive
  inertial confinement fusion},}\ }\href {\doibase 10.1063/1.4985150}
  {\bibfield  {journal} {\bibinfo  {journal} {Physics of Plasmas}\ }\textbf
  {\bibinfo {volume} {24}},\ \bibinfo {pages} {062708} (\bibinfo {year}
  {2017})}\BibitemShut {NoStop}%
\bibitem [{\citenamefont {Walsh}\ \emph {et~al.}(2019)\citenamefont {Walsh},
  \citenamefont {McGlinchey}, \citenamefont {Tong}, \citenamefont {Appelbe},
  \citenamefont {Crilly}, \citenamefont {Zhang},\ and\ \citenamefont
  {Chittenden}}]{walsh2019}%
  \BibitemOpen
  \bibfield  {author} {\bibinfo {author} {\bibfnamefont {C.~A.}\ \bibnamefont
  {Walsh}}, \bibinfo {author} {\bibfnamefont {K.}~\bibnamefont {McGlinchey}},
  \bibinfo {author} {\bibfnamefont {J.~K.}\ \bibnamefont {Tong}}, \bibinfo
  {author} {\bibfnamefont {B.~D.}\ \bibnamefont {Appelbe}}, \bibinfo {author}
  {\bibfnamefont {A.}~\bibnamefont {Crilly}}, \bibinfo {author} {\bibfnamefont
  {M.~F.}\ \bibnamefont {Zhang}}, \ and\ \bibinfo {author} {\bibfnamefont
  {J.~P.}\ \bibnamefont {Chittenden}},\ }\bibfield  {title} {\enquote {\bibinfo
  {title} {Perturbation modifications by pre-magnetisation of inertial
  confinement fusion implosions},}\ }\href {\doibase 10.1063/1.5085498}
  {\bibfield  {journal} {\bibinfo  {journal} {Physics of Plasmas}\ }\textbf
  {\bibinfo {volume} {26}},\ \bibinfo {pages} {022701} (\bibinfo {year}
  {2019})}\BibitemShut {NoStop}%
\bibitem [{\citenamefont {Slutz}\ and\ \citenamefont
  {Vesey}(2012)}]{slutz2012}%
  \BibitemOpen
  \bibfield  {author} {\bibinfo {author} {\bibfnamefont {S.~A.}\ \bibnamefont
  {Slutz}}\ and\ \bibinfo {author} {\bibfnamefont {R.~A.}\ \bibnamefont
  {Vesey}},\ }\bibfield  {title} {\enquote {\bibinfo {title} {High-{{Gain
  Magnetized Inertial Fusion}}},}\ }\href {\doibase
  10.1103/PhysRevLett.108.025003} {\bibfield  {journal} {\bibinfo  {journal}
  {Phys. Rev. Lett.}\ }\textbf {\bibinfo {volume} {108}},\ \bibinfo {pages}
  {025003} (\bibinfo {year} {2012})}\BibitemShut {NoStop}%
\bibitem [{\citenamefont {Sefkow}\ \emph {et~al.}(2014)\citenamefont {Sefkow},
  \citenamefont {Slutz}, \citenamefont {Koning}, \citenamefont {Marinak},
  \citenamefont {Peterson}, \citenamefont {Sinars},\ and\ \citenamefont
  {Vesey}}]{sefkow2014}%
  \BibitemOpen
  \bibfield  {author} {\bibinfo {author} {\bibfnamefont {A.~B.}\ \bibnamefont
  {Sefkow}}, \bibinfo {author} {\bibfnamefont {S.~A.}\ \bibnamefont {Slutz}},
  \bibinfo {author} {\bibfnamefont {J.~M.}\ \bibnamefont {Koning}}, \bibinfo
  {author} {\bibfnamefont {M.~M.}\ \bibnamefont {Marinak}}, \bibinfo {author}
  {\bibfnamefont {K.~J.}\ \bibnamefont {Peterson}}, \bibinfo {author}
  {\bibfnamefont {D.~B.}\ \bibnamefont {Sinars}}, \ and\ \bibinfo {author}
  {\bibfnamefont {R.~A.}\ \bibnamefont {Vesey}},\ }\bibfield  {title}
  {{ \enquote {\bibinfo {title} {Design of magnetized liner
  inertial fusion experiments using the {{Z}} facility},}\ }}\href {\doibase
  10.1063/1.4890298} {\bibfield  {journal} {\bibinfo  {journal} {Physics of
  Plasmas}\ }\textbf {\bibinfo {volume} {21}},\ \bibinfo {pages} {072711}
  (\bibinfo {year} {2014})}\BibitemShut {NoStop}%
\bibitem [{\citenamefont {Chang}\ \emph {et~al.}(2011)\citenamefont {Chang},
  \citenamefont {Fiksel}, \citenamefont {Hohenberger}, \citenamefont {Knauer},
  \citenamefont {Betti}, \citenamefont {Marshall}, \citenamefont {Meyerhofer},
  \citenamefont {S{\'e}guin},\ and\ \citenamefont {Petrasso}}]{chang2011}%
  \BibitemOpen
  \bibfield  {author} {\bibinfo {author} {\bibfnamefont {P.~Y.}\ \bibnamefont
  {Chang}}, \bibinfo {author} {\bibfnamefont {G.}~\bibnamefont {Fiksel}},
  \bibinfo {author} {\bibfnamefont {M.}~\bibnamefont {Hohenberger}}, \bibinfo
  {author} {\bibfnamefont {J.~P.}\ \bibnamefont {Knauer}}, \bibinfo {author}
  {\bibfnamefont {R.}~\bibnamefont {Betti}}, \bibinfo {author} {\bibfnamefont
  {F.~J.}\ \bibnamefont {Marshall}}, \bibinfo {author} {\bibfnamefont {D.~D.}\
  \bibnamefont {Meyerhofer}}, \bibinfo {author} {\bibfnamefont {F.~H.}\
  \bibnamefont {S{\'e}guin}}, \ and\ \bibinfo {author} {\bibfnamefont {R.~D.}\
  \bibnamefont {Petrasso}},\ }\bibfield  {title} {\enquote {\bibinfo {title}
  {Fusion {{Yield Enhancement}} in {{Magnetized Laser}}-{{Driven
  Implosions}}},}\ }\href {\doibase 10.1103/PhysRevLett.107.035006} {\bibfield
  {journal} {\bibinfo  {journal} {Phys. Rev. Lett.}\ }\textbf {\bibinfo
  {volume} {107}},\ \bibinfo {pages} {035006} (\bibinfo {year}
  {2011})}\BibitemShut {NoStop}%
\bibitem [{\citenamefont {Hohenberger}\ \emph {et~al.}(2012)\citenamefont
  {Hohenberger}, \citenamefont {Chang}, \citenamefont {Fiksel}, \citenamefont
  {Knauer}, \citenamefont {Betti}, \citenamefont {Marshall}, \citenamefont
  {Meyerhofer}, \citenamefont {S{\'e}guin},\ and\ \citenamefont
  {Petrasso}}]{hohenberger2012}%
  \BibitemOpen
  \bibfield  {author} {\bibinfo {author} {\bibfnamefont {M.}~\bibnamefont
  {Hohenberger}}, \bibinfo {author} {\bibfnamefont {P.-Y.}\ \bibnamefont
  {Chang}}, \bibinfo {author} {\bibfnamefont {G.}~\bibnamefont {Fiksel}},
  \bibinfo {author} {\bibfnamefont {J.~P.}\ \bibnamefont {Knauer}}, \bibinfo
  {author} {\bibfnamefont {R.}~\bibnamefont {Betti}}, \bibinfo {author}
  {\bibfnamefont {F.~J.}\ \bibnamefont {Marshall}}, \bibinfo {author}
  {\bibfnamefont {D.~D.}\ \bibnamefont {Meyerhofer}}, \bibinfo {author}
  {\bibfnamefont {F.~H.}\ \bibnamefont {S{\'e}guin}}, \ and\ \bibinfo {author}
  {\bibfnamefont {R.~D.}\ \bibnamefont {Petrasso}},\ }\bibfield  {title}
  {\enquote {\bibinfo {title} {Inertial confinement fusion implosions with
  imposed magnetic field compression using the {{OMEGA Laser}}},}\ }\href
  {\doibase 10.1063/1.3696032} {\bibfield  {journal} {\bibinfo  {journal}
  {Physics of Plasmas}\ }\textbf {\bibinfo {volume} {19}},\ \bibinfo {pages}
  {056306} (\bibinfo {year} {2012})}\BibitemShut {NoStop}%
\bibitem [{\citenamefont {Manuel}\ \emph {et~al.}(2015)\citenamefont {Manuel},
  \citenamefont {Kuranz}, \citenamefont {Rasmus}, \citenamefont {Klein},
  \citenamefont {MacDonald}, \citenamefont {Trantham}, \citenamefont {Fein},
  \citenamefont {Belancourt}, \citenamefont {Young}, \citenamefont {Keiter},
  \citenamefont {Drake}, \citenamefont {Pollock}, \citenamefont {Park},
  \citenamefont {Hazi}, \citenamefont {Williams},\ and\ \citenamefont
  {Chen}}]{manuel2015}%
  \BibitemOpen
  \bibfield  {author} {\bibinfo {author} {\bibfnamefont {M.~J.~E.}\
  \bibnamefont {Manuel}}, \bibinfo {author} {\bibfnamefont {C.~C.}\
  \bibnamefont {Kuranz}}, \bibinfo {author} {\bibfnamefont {A.~M.}\
  \bibnamefont {Rasmus}}, \bibinfo {author} {\bibfnamefont {S.~R.}\
  \bibnamefont {Klein}}, \bibinfo {author} {\bibfnamefont {M.~J.}\ \bibnamefont
  {MacDonald}}, \bibinfo {author} {\bibfnamefont {M.~R.}\ \bibnamefont
  {Trantham}}, \bibinfo {author} {\bibfnamefont {J.~R.}\ \bibnamefont {Fein}},
  \bibinfo {author} {\bibfnamefont {P.~X.}\ \bibnamefont {Belancourt}},
  \bibinfo {author} {\bibfnamefont {R.~P.}\ \bibnamefont {Young}}, \bibinfo
  {author} {\bibfnamefont {P.~A.}\ \bibnamefont {Keiter}}, \bibinfo {author}
  {\bibfnamefont {R.~P.}\ \bibnamefont {Drake}}, \bibinfo {author}
  {\bibfnamefont {B.~B.}\ \bibnamefont {Pollock}}, \bibinfo {author}
  {\bibfnamefont {J.}~\bibnamefont {Park}}, \bibinfo {author} {\bibfnamefont
  {A.~U.}\ \bibnamefont {Hazi}}, \bibinfo {author} {\bibfnamefont {G.~J.}\
  \bibnamefont {Williams}}, \ and\ \bibinfo {author} {\bibfnamefont
  {H.}~\bibnamefont {Chen}},\ }\bibfield  {title} {{
  \enquote {\bibinfo {title} {Experimental results from magnetized-jet
  experiments executed at the {{Jupiter Laser Facility}}},}\ }}\href {\doibase
  10.1016/j.hedp.2014.07.003} {\bibfield  {journal} {\bibinfo  {journal} {High
  Energy Density Physics}\ }\bibinfo {series} {10th {{International
  Conference}} on {{High Energy Density Laboratory Astrophysics}}},\ \textbf
  {\bibinfo {volume} {17}},\ \bibinfo {pages} {52--62} (\bibinfo {year}
  {2015})}\BibitemShut {NoStop}%
\bibitem [{\citenamefont {Albertazzi}\ \emph {et~al.}(18ed)\citenamefont
  {Albertazzi}, \citenamefont {Falize}, \citenamefont {Pelka}, \citenamefont
  {Brack}, \citenamefont {Kroll}, \citenamefont {Yurchak}, \citenamefont
  {Brambrink}, \citenamefont {Mabey}, \citenamefont {Ozaki}, \citenamefont
  {Pikuz}, \citenamefont {Som}, \citenamefont {{Bonnet-Bidaud}}, \citenamefont
  {Cross}, \citenamefont {Filippov}, \citenamefont {Gregori}, \citenamefont
  {Kodama}, \citenamefont {Mouchet}, \citenamefont {Morita}, \citenamefont
  {Sakawa}, \citenamefont {Drake}, \citenamefont {Kuranz}, \citenamefont
  {Manuel}, \citenamefont {Li}, \citenamefont {Tzeferacos}, \citenamefont
  {Lamb}, \citenamefont {Schramm},\ and\ \citenamefont
  {Koenig}}]{albertazzi2018}%
  \BibitemOpen
  \bibfield  {author} {\bibinfo {author} {\bibfnamefont {B.}~\bibnamefont
  {Albertazzi}}, \bibinfo {author} {\bibfnamefont {E.}~\bibnamefont {Falize}},
  \bibinfo {author} {\bibfnamefont {A.}~\bibnamefont {Pelka}}, \bibinfo
  {author} {\bibfnamefont {F.}~\bibnamefont {Brack}}, \bibinfo {author}
  {\bibfnamefont {F.}~\bibnamefont {Kroll}}, \bibinfo {author} {\bibfnamefont
  {R.}~\bibnamefont {Yurchak}}, \bibinfo {author} {\bibfnamefont
  {E.}~\bibnamefont {Brambrink}}, \bibinfo {author} {\bibfnamefont
  {P.}~\bibnamefont {Mabey}}, \bibinfo {author} {\bibfnamefont
  {N.}~\bibnamefont {Ozaki}}, \bibinfo {author} {\bibfnamefont
  {S.}~\bibnamefont {Pikuz}}, \bibinfo {author} {\bibfnamefont {L.~V.~B.}\
  \bibnamefont {Som}}, \bibinfo {author} {\bibfnamefont {J.~M.}\ \bibnamefont
  {{Bonnet-Bidaud}}}, \bibinfo {author} {\bibfnamefont {J.~E.}\ \bibnamefont
  {Cross}}, \bibinfo {author} {\bibfnamefont {E.}~\bibnamefont {Filippov}},
  \bibinfo {author} {\bibfnamefont {G.}~\bibnamefont {Gregori}}, \bibinfo
  {author} {\bibfnamefont {R.}~\bibnamefont {Kodama}}, \bibinfo {author}
  {\bibfnamefont {M.}~\bibnamefont {Mouchet}}, \bibinfo {author} {\bibfnamefont
  {T.}~\bibnamefont {Morita}}, \bibinfo {author} {\bibfnamefont
  {Y.}~\bibnamefont {Sakawa}}, \bibinfo {author} {\bibfnamefont {R.~P.}\
  \bibnamefont {Drake}}, \bibinfo {author} {\bibfnamefont {C.~C.}\ \bibnamefont
  {Kuranz}}, \bibinfo {author} {\bibfnamefont {M.~J.-E.}\ \bibnamefont
  {Manuel}}, \bibinfo {author} {\bibfnamefont {C.}~\bibnamefont {Li}}, \bibinfo
  {author} {\bibfnamefont {P.}~\bibnamefont {Tzeferacos}}, \bibinfo {author}
  {\bibfnamefont {D.}~\bibnamefont {Lamb}}, \bibinfo {author} {\bibfnamefont
  {U.}~\bibnamefont {Schramm}}, \ and\ \bibinfo {author} {\bibfnamefont
  {M.}~\bibnamefont {Koenig}},\ }\bibfield  {title} {{
  \enquote {\bibinfo {title} {Experimental platform for the investigation
  of magnetized-reverse-shock dynamics in the context of {{POLAR}}},}\ }}\href
  {\doibase 10.1017/hpl.2018.37} {\bibfield  {journal} {\bibinfo  {journal}
  {High Power Laser Sci. Eng.}\ }\textbf {\bibinfo {volume} {6}} (\bibinfo
  {year} {2018/ed}),\ 10.1017/hpl.2018.37}\BibitemShut {NoStop}%
\bibitem [{\citenamefont {Bradford}\ \emph {et~al.}(20ed)\citenamefont
  {Bradford}, \citenamefont {Read}, \citenamefont {Ehret}, \citenamefont
  {Antonelli}, \citenamefont {Khan}, \citenamefont {Booth}, \citenamefont
  {Glize}, \citenamefont {Carroll}, \citenamefont {Clarke}, \citenamefont
  {Heathcote}, \citenamefont {Ryazantsev}, \citenamefont {Pikuz}, \citenamefont
  {Spindloe}, \citenamefont {Moody}, \citenamefont {Pollock}, \citenamefont
  {Tikhonchuk}, \citenamefont {Ridgers}, \citenamefont {Santos},\ and\
  \citenamefont {Woolsey}}]{bradford2020}%
  \BibitemOpen
  \bibfield  {author} {\bibinfo {author} {\bibfnamefont {P.}~\bibnamefont
  {Bradford}}, \bibinfo {author} {\bibfnamefont {M.~P.}\ \bibnamefont {Read}},
  \bibinfo {author} {\bibfnamefont {M.}~\bibnamefont {Ehret}}, \bibinfo
  {author} {\bibfnamefont {L.}~\bibnamefont {Antonelli}}, \bibinfo {author}
  {\bibfnamefont {M.}~\bibnamefont {Khan}}, \bibinfo {author} {\bibfnamefont
  {N.}~\bibnamefont {Booth}}, \bibinfo {author} {\bibfnamefont
  {K.}~\bibnamefont {Glize}}, \bibinfo {author} {\bibfnamefont
  {D.}~\bibnamefont {Carroll}}, \bibinfo {author} {\bibfnamefont {R.~J.}\
  \bibnamefont {Clarke}}, \bibinfo {author} {\bibfnamefont {R.}~\bibnamefont
  {Heathcote}}, \bibinfo {author} {\bibfnamefont {S.}~\bibnamefont
  {Ryazantsev}}, \bibinfo {author} {\bibfnamefont {S.}~\bibnamefont {Pikuz}},
  \bibinfo {author} {\bibfnamefont {C.}~\bibnamefont {Spindloe}}, \bibinfo
  {author} {\bibfnamefont {J.~D.}\ \bibnamefont {Moody}}, \bibinfo {author}
  {\bibfnamefont {B.~B.}\ \bibnamefont {Pollock}}, \bibinfo {author}
  {\bibfnamefont {V.~T.}\ \bibnamefont {Tikhonchuk}}, \bibinfo {author}
  {\bibfnamefont {C.~P.}\ \bibnamefont {Ridgers}}, \bibinfo {author}
  {\bibfnamefont {J.~J.}\ \bibnamefont {Santos}}, \ and\ \bibinfo {author}
  {\bibfnamefont {N.~C.}\ \bibnamefont {Woolsey}},\ }\bibfield  {title}
  {{ \enquote {\bibinfo {title} {Proton deflectometry of a
  capacitor coil target along two axes},}\ }}\href {\doibase
  10.1017/hpl.2020.9} {\bibfield  {journal} {\bibinfo  {journal} {High Power
  Laser Sci. Eng.}\ }\textbf {\bibinfo {volume} {8}} (\bibinfo {year}
  {2020/ed}),\ 10.1017/hpl.2020.9}\BibitemShut {NoStop}%
\bibitem [{\citenamefont {Arber}\ \emph {et~al.}(2015)\citenamefont {Arber},
  \citenamefont {Bennett}, \citenamefont {Brady}, \citenamefont
  {{Lawrence-Douglas}}, \citenamefont {Ramsay}, \citenamefont {Sircombe},
  \citenamefont {Gillies}, \citenamefont {Evans}, \citenamefont {Schmitz},
  \citenamefont {Bell},\ and\ \citenamefont {Ridgers}}]{arber2015}%
  \BibitemOpen
  \bibfield  {author} {\bibinfo {author} {\bibfnamefont {T.~D.}\ \bibnamefont
  {Arber}}, \bibinfo {author} {\bibfnamefont {K.}~\bibnamefont {Bennett}},
  \bibinfo {author} {\bibfnamefont {C.~S.}\ \bibnamefont {Brady}}, \bibinfo
  {author} {\bibfnamefont {A.}~\bibnamefont {{Lawrence-Douglas}}}, \bibinfo
  {author} {\bibfnamefont {M.~G.}\ \bibnamefont {Ramsay}}, \bibinfo {author}
  {\bibfnamefont {N.~J.}\ \bibnamefont {Sircombe}}, \bibinfo {author}
  {\bibfnamefont {P.}~\bibnamefont {Gillies}}, \bibinfo {author} {\bibfnamefont
  {R.~G.}\ \bibnamefont {Evans}}, \bibinfo {author} {\bibfnamefont
  {H.}~\bibnamefont {Schmitz}}, \bibinfo {author} {\bibfnamefont {A.~R.}\
  \bibnamefont {Bell}}, \ and\ \bibinfo {author} {\bibfnamefont {C.~P.}\
  \bibnamefont {Ridgers}},\ }\bibfield  {title} {\enquote {\bibinfo {title}
  {Contemporary particle-in-cell approach to laser-plasma modelling},}\ }\href
  {\doibase 10.1088/0741-3335/57/11/113001} {\bibfield  {journal} {\bibinfo
  {journal} {Plasma Phys Control Fusion}\ }\textbf {\bibinfo {volume} {57}},\
  \bibinfo {pages} {113001} (\bibinfo {year} {2015})}\BibitemShut {NoStop}%
\bibitem [{\citenamefont {Richardson}(1972)}]{richardson1972}%
  \BibitemOpen
  \bibfield  {author} {\bibinfo {author} {\bibfnamefont {W.~H.}\ \bibnamefont
  {Richardson}},\ }\bibfield  {title} {{ \enquote {\bibinfo
  {title} {Bayesian-{{Based Iterative Method}} of {{Image Restoration}}*},}\
  }}\href {\doibase 10.1364/JOSA.62.000055} {\bibfield  {journal} {\bibinfo
  {journal} {J. Opt. Soc. Am., JOSA}\ }\textbf {\bibinfo {volume} {62}},\
  \bibinfo {pages} {55--59} (\bibinfo {year} {1972})}\BibitemShut {NoStop}%
\bibitem [{\citenamefont {Lucy}(1974)}]{lucy1974}%
  \BibitemOpen
  \bibfield  {author} {\bibinfo {author} {\bibfnamefont {L.~B.}\ \bibnamefont
  {Lucy}},\ }\bibfield  {title} {\enquote {\bibinfo {title} {An iterative
  technique for the rectification of observed distributions},}\ }\href
  {\doibase 10.1086/111605} {\bibfield  {journal} {\bibinfo  {journal} {The
  Astronomical Journal}\ }\textbf {\bibinfo {volume} {79}},\ \bibinfo {pages}
  {745} (\bibinfo {year} {1974})}\BibitemShut {NoStop}%
\end{thebibliography}
%

\end{document}